\documentclass[aps,prc,reprint,nofootinbib,superscriptaddress,floatfix]{revtex4-2}
\usepackage{amsmath,amssymb,bm}
\usepackage{graphicx}
\usepackage{booktabs}
\usepackage{siunitx}
\usepackage{microtype}
\usepackage[colorlinks=true,linkcolor=blue,citecolor=blue,urlcolor=blue]{hyperref}
\begin{document}

\title{Real-time decay dynamics of the low-lying $1/2^+$ state in $^9$Be: Signatures of virtual-state behavior}

\author{Yuma Kikuchi}
\email{kikuchi@tokuyama.ac.jp}
\affiliation{National Institute of Technology, Tokuyama College, Shunan 745-8585, Japan}
\affiliation{RIKEN Nishina Center, 2-1 Hirosawa, Wako 351-0198, Japan}

\author{Kiyoshi Kat\={o}}
\affiliation{Nuclear Reaction Data Centre, Faculty of Science, Hokkaido University, Sapporo 060-0810, Japan}

\author{Takayuki Myo}
\affiliation{General Education, Faculty of Engineering, Osaka Institute of Technology, Osaka 535-8585, Japan}
\affiliation{Research Center for Nuclear Physics (RCNP), Osaka University, Ibaraki 567-0047, Japan}

\date{\today}

\begin{abstract}
\textbf{Background:} The low-lying $1/2^+$ state of $^9$Be has been interpreted either as a near-threshold resonance or as an $^8$Be+$n$ virtual state. Because both pictures can produce a pronounced threshold enhancement, its nature is difficult to determine from energy-domain observables alone.

\textbf{Purpose:} We investigate the low-lying $1/2^+$ state through its real-time decay dynamics and compare its behavior with that of the well-established $5/2^+$ resonance.

\textbf{Methods:} A wave packet produced by electric-dipole excitation of the $^9$Be ground state is propagated in an $\alpha+\alpha+n$ three-body model. We analyze the survival probability, the fit-window dependence of the effective decay width, and time-dependent densities in different Jacobi-coordinate representations. The $1/2^+$ survival dynamics are further compared with an $s$-wave effective-range description motivated by the dominant $^8$Be+$n$ component found in our previous three-body studies.

\textbf{Results:} The $5/2^+$ channel shows approximately exponential late-time decay, with an effective width of $0.264$ MeV, close to the complex-scaled pole width of $0.285$ MeV. In contrast, the $1/2^+$ effective width decreases as the fitting interval is shifted to later times, and no stationary decay width is obtained. Its late-time survival probability is well described by the effective-range spectral distribution. The density evolution develops an $^8$Be+$n$-like asymptotic configuration while retaining a localized $p_{3/2}$ $\alpha n$ component suggestive of a $^5$He+$\alpha$-like configuration in the internal region.

\textbf{Conclusions:} The $1/2^+$ decay differs qualitatively from resonance-like exponential decay and is consistent with near-threshold virtual-state dynamics. The distinct internal and asymptotic configurations revealed by the time-dependent densities provide complementary structural information beyond that contained in a single decay width.
\end{abstract}

\maketitle

\section{Introduction}

Nuclear states near particle-emission thresholds are strongly influenced by coupling to the continuum~\cite{Michel2010,Johnson2020}. In this region, different continuum structures can produce similar enhancements in energy-domain observables, making the underlying nature of a threshold state difficult to identify from spectra alone. The distinction between resonances and virtual states is a particularly important example. Although they correspond to different analytic structures of the scattering amplitude, both can generate pronounced near-threshold strength. Their decay dynamics may therefore provide information not directly accessible from the energy-domain response.

The low-lying $1/2^+$ state of $^9$Be is a representative case. Its near-threshold photodisintegration strength has been studied experimentally over many decades, from early radioactive-source and bremsstrahlung measurements~\cite{Gibbons1959,Jakobson1961,John1962,Berman1967} to laser-Compton and quasi-monochromatic photon-beam measurements~\cite{Utsunomiya2001,Arnold2012,Utsunomiya2015}. Effective-range analyses describe the enhancement in terms of an $s$-wave $^8$Be+$n$ virtual state with a pole only a few tens of keV below threshold~\cite{EfrosBang1999}. In contrast, Garrido, Fedorov, and Jensen proposed a genuine three-body resonance~\cite{Garrido2010}, in which the configuration evolves from a short-distance $\alpha+{}^5$He-like structure to a large-distance $n+{}^8$Be-like structure and the associated rearrangement generates an effective barrier. Resonance-like parametrizations have also been discussed in analyses of the electromagnetic response~\cite{Efros2014}. Previous $\alpha+\alpha+n$ complex-scaling calculations reproduced the low-energy photodisintegration strength without identifying an isolated $1/2^+$ resonance pole over the scaling angles considered and emphasized the role of the $^8$Be+$n$ continuum~\cite{Odsuren2015,Kikuchi2016}. These contrasting descriptions motivate a dynamical examination of the state.

In the present work, we study the real-time decay of a wave packet produced by electric-dipole excitation of the $^9$Be ground state. We employ the complex-scaled time-evolution formalism developed in our previous work~\cite{Kikuchi2026Time}, based on the complex scaling method~\cite{Aguilar1971,Balslev1971,Aoyama2006}. This framework treats continuum-wave-packet propagation within the same complex-scaled basis representation used for resonance and continuum calculations. Within an $\alpha+\alpha+n$ three-body model, we analyze the survival probability and the dependence of the extracted effective width on the fitting interval. The well-defined $5/2^+$ resonance~\cite{Tilley2004} serves as a reference for decay governed by an identifiable resonance pole. For the $1/2^+$ channel, we test whether the late-time behavior approaches a stationary exponential decay and compare the calculated survival probability with an effective-range description of an $s$-wave virtual state.

We also examine time-dependent spatial densities in different Jacobi coordinate representations. This allows the internal $\alpha n$ correlation and the outward-developing $^8$Be+$n$-like geometry to be followed within the same wave packet, complementing the survival analysis with spatial information. By combining the temporal and spatial information, we characterize the $1/2^+$ continuum without assuming that it can be represented by a single resonance width.

The remainder of this paper is organized as follows. Section~II describes the three-body model, time-evolution framework, and spatial densities. Section~III introduces the effective-range description. Section~IV presents the decay dynamics and time-dependent spatial structure, and Sec.~V summarizes the conclusions.

\section{Three-body model and time evolution}

The calculation follows the complex-scaled time-evolution formalism developed in Ref.~\cite{Kikuchi2026Time}. Its implementation for explicit three-body decay, including rearranged Jacobi-coordinate bases, is described in our recent $^6$Be study~\cite{Kikuchi6Be2026}. Here we summarize the ingredients specific to the present $^9$Be calculation and the observables used below.

\subsection{$\alpha+\alpha+n$ model}

We describe $^9$Be by the three-body Hamiltonian
\begin{equation}
H=T+V_{\alpha n}^{(1)}+V_{\alpha n}^{(2)}+V_{\alpha\alpha}
  +V_{\rm PF}+V_3,
\label{eq:H}
\end{equation}
where $T$ is the three-body kinetic-energy operator with the center-of-mass motion removed, $V_{\alpha n}^{(1)}$ and $V_{\alpha n}^{(2)}$ are the interactions between the neutron and the two $\alpha$ particles, and $V_{\alpha\alpha}$ is the $\alpha$--$\alpha$ interaction. The $\alpha$--$n$ interaction is the KKNN interaction~\cite{Kanada1979}. The $\alpha$--$\alpha$ interaction, including the folded Coulomb term, is the same as in our previous $^9$Be calculations~\cite{Odsuren2015,Kikuchi2016}. Pauli-forbidden components are removed with the orthogonality-condition-model pseudopotential $V_{\rm PF}$~\cite{Krasnopolsky1975}. The state-dependent three-body interaction has the same form as in the previous photodisintegration calculation~\cite{Kikuchi2016},
\begin{equation}
 V_3
 =v_3\exp(-\mu_3\rho^2),
 \qquad
 \rho^2
 =2r_{\alpha\alpha}^2
 +\frac{8}{9}R_{n-(\alpha\alpha)}^2,
\label{eq:V3}
\end{equation}
where $v_3$ is adjusted for each spin-parity state, while the common range parameter is $\mu_3=0.02~{\rm fm}^{-2}$.

The wave functions are expanded using the Gaussian expansion method (GEM)~\cite{Hiyama2003} for the two rearrangement channels, $(\alpha n)+\alpha$ and $(\alpha\alpha)+n$. Including both Jacobi coordinate sets is essential for describing, within the same three-body wave function, compact $\alpha n$ correlations and the $n+(\alpha\alpha)$ asymptotic configuration discussed below. Schematically, the three-body wave function is written as
\begin{equation}
 \Psi_{JM}
 =\sum_{c}\sum_{\beta} C^{(c)}_{\beta}\,
 \mathcal{S}_{\alpha}
 \left[
  \left[
   \phi^{(c)}_{nl}(\bm r_c)\,
   \varphi^{(c)}_{NL}(\bm R_c)
  \right]_{\Lambda}
  \chi_{1/2}
 \right]_{JM},
\label{eq:GEMexpansion}
\end{equation}
where $c$ labels the Jacobi rearrangement and $C^{(c)}_{\beta}$ is the corresponding expansion coefficient. We use the collective index $\beta=(n,l,N,L,\Lambda)$, where $n$ and $N$ label the Gaussian ranges, $l$ and $L$ are the orbital angular momenta associated with $\bm r_c$ and $\bm R_c$, and $\Lambda$ is their coupled orbital angular momentum. The operator $\mathcal{S}_{\alpha}$ symmetrizes the wave function under exchange of the two $\alpha$ particles. The magnetic substates of the Gaussian basis functions are written with solid spherical harmonics as
\begin{align}
 \phi^{(c)}_{nlm}(\bm r_c)
 &=\mathcal N_{nl}\,e^{-(r_c/r_n)^2}\mathcal{Y}_{lm}(\bm r_c),\\
 \varphi^{(c)}_{NLM_L}(\bm R_c)
 &=\mathcal N_{NL}\,e^{-(R_c/R_N)^2}\mathcal{Y}_{LM_L}(\bm R_c),
\label{eq:GaussianBasis}
\end{align}
where $\mathcal N_{nl}$ and $\mathcal N_{NL}$ are normalization constants, and $r_n$ and $R_N$ are the Gaussian ranges. The Gaussian ranges are distributed geometrically. For the $3/2^-$ ground state, 15 Gaussian ranges are used for each relative coordinate over $0.10$--$15$ fm. For the positive-parity final states, 25 Gaussian ranges are used for each relative coordinate over $0.15$--$50$ fm.

\begin{table}[t]
\caption{Three-body-force strengths and calculated and experimental energies and widths for the low-lying states considered here. Energies are measured from the $\alpha+\alpha+n$ threshold, and all energies and widths are given in MeV. Experimental values are taken from Ref.~\cite{Tilley2004}.}
\label{tab:v3_parameters}
\begin{ruledtabular}
\begin{tabular}{cccc}
$J^\pi$ &
$v_3$ (MeV) &
$(E,\Gamma)_{\rm calc}$ &
$(E,\Gamma)_{\rm exp}$ \\
\hline
$3/2^-$ & $1.10$  & $(-1.570,\;-)$ & $(-1.574,\;-)$ \\
$1/2^+$ & $-0.90$ & $(-,\;-)^{a}$ & $(0.111,\;0.217)$ \\
$5/2^+$ & $-0.30$ & $(1.394,\;0.285)$ & $(1.475,\;0.282)$ \\
\end{tabular}
\end{ruledtabular}
\begin{flushleft}
\footnotesize
$^{a}$ No isolated $1/2^+$ pole is obtained in the complex-scaled spectrum.
\end{flushleft}
\end{table}

\subsection{Complex-scaled time evolution and decay observables}

Under the complex-scaling transformation,
\begin{equation}
 \bm r\rightarrow \bm r e^{i\theta},
 \qquad
 \bm R\rightarrow \bm R e^{i\theta},
\end{equation}
where $\theta$ is the scaling angle. In the present calculation, we use $\theta=10^\circ$. With $\hat U(\theta)$ denoting the scaling operator, the complex-scaled Hamiltonian is
\begin{equation}
 H^\theta=\hat U(\theta)H\hat U^{-1}(\theta).
\end{equation}
The complex-scaled Schr\"odinger equation is
\begin{equation}
 H^\theta|\Psi_\nu^\theta\rangle
 =E_\nu^\theta|\Psi_\nu^\theta\rangle,
\label{eq:CSM_Schrodinger}
\end{equation}
where $E_\nu^\theta$ and $|\Psi_\nu^\theta\rangle$ are the complex eigenvalue and corresponding eigenstate, respectively. The state biorthogonal to $|\Psi_\nu^\theta\rangle$ is denoted by $\langle\widetilde{\Psi}_\nu^\theta|$. In the finite $L^2$ model space, these states satisfy the extended completeness relation
\begin{equation}
 \sum_\nu
 |\Psi_\nu^\theta\rangle
 \langle\widetilde{\Psi}_\nu^\theta|
 =1^\theta,
\label{eq:ECR}
\end{equation}
where $1^\theta$ denotes the identity operator in the complex-scaled model space. The sum contains the bound, resonant, and discretized nonresonant continuum states retained in the calculation. In this representation, resonant poles are separated from the nonresonant continuum, while the continuum spectrum is rotated into the complex-energy plane~\cite{Aguilar1971,Balslev1971,Aoyama2006}.

The initial positive-parity wave packet $|\Phi(0)\rangle$ is generated by electric-dipole excitation of the ground-state wave function $|\Psi_{\rm gs}\rangle$,
\begin{equation}
 |\Phi(0)\rangle \propto \hat M(E1)|\Psi_{\rm gs}\rangle .
\label{eq:initial_packet}
\end{equation}
The electric-dipole operator has spherical components
\begin{equation}
 \hat M_{1\mu}(E1)
 =e\sum_i Z_i\,
 \mathcal{Y}_{1\mu}(\bm R_i-\bm R_{\rm cm}),
\label{eq:E1op}
\end{equation}
where $Z_i$ and $\bm R_i$ are the charge number and coordinate of the $i$th constituent, respectively, and $\bm R_{\rm cm}$ is the center-of-mass coordinate. For the $\alpha+\alpha+n$ system, this operator excites the relative motion of the neutron against the charged $\alpha\alpha$ subsystem.

Using the extended completeness relation in Eq.~(\ref{eq:ECR}), the time-evolution operator is written as
\begin{equation}
 e^{-iHt/\hbar}
 =
 \hat U^{-1}(\theta)
 \sum_\nu
 e^{-iE_\nu^\theta t/\hbar}
 |\Psi_\nu^\theta\rangle
 \langle\widetilde{\Psi}_\nu^\theta|
 \hat U(\theta),
\end{equation}
and the wave packet therefore evolves as~\cite{Kikuchi2026Time,Kikuchi6Be2026}
\begin{equation}
 |\Phi(t)\rangle
 =\hat U^{-1}(\theta)
 \sum_{\nu} e^{-iE_\nu^\theta t/\hbar}
 |\Psi_\nu^\theta\rangle
 \langle\widetilde{\Psi}_\nu^\theta|
 \hat U(\theta)|\Phi(0)\rangle .
\label{eq:csm_time}
\end{equation}
Thus, resonant and nonresonant continuum components contained in the initial $E1$ wave packet are propagated within the same spectral representation. Details of the derivation and numerical implementation are given in Refs.~\cite{Kikuchi2026Time,Kikuchi6Be2026}.

The normalized survival amplitude and probability are
\begin{align}
 A(t)&=\frac{\langle\Phi(0)|\Phi(t)\rangle}
 {\langle\Phi(0)|\Phi(0)\rangle},\\
 P_{\rm surv}(t)&=|A(t)|^2 .
\label{eq:survival}
\end{align}
To characterize the decay without assuming a line shape, we extract an effective width from a linear fit to the logarithmic survival probability over a finite time interval,
\begin{equation}
 -\ln P_{\rm surv}(t)\simeq C+\frac{\Gamma_{\rm eff}}{\hbar}t,
 \qquad t_{\min}\le t\le t_{\max},
\label{eq:gamma_local}
\end{equation}
where $\Gamma_{\rm eff}$ is the effective decay width and $t_{\min}\le t\le t_{\max}$ is the fitting interval. For decay dominated by a single resonance pole, $\Gamma_{\rm eff}$ should become approximately insensitive to the fitting window, as found for the $^6$Be three-body decay in Ref.~\cite{Kikuchi6Be2026}. We therefore use the fit-window dependence of $\Gamma_{\rm eff}$ as a diagnostic of the late-time decay dynamics.

\subsection{Time-dependent spatial density}

To examine the spatial development of the $1/2^+$ wave packet, we evaluate the two-dimensional radial density in each Jacobi coordinate set,
\begin{equation}
 D^{(c)}(r,R;t)=r^2R^2\int d\Omega_r\,d\Omega_R
 \left|\Phi^{(c)}(\bm r,\bm R;t)\right|^2,
\label{eq:density}
\end{equation}
where $c$ denotes the Jacobi representation and $\Phi^{(c)}(\bm r,\bm R;t)$ is the wave packet expressed in that coordinate set. With the angular variables integrated out, the density is normalized through
\begin{equation}
 \int_0^\infty dr\int_0^\infty dR\,
 D^{(c)}(r,R;t)
 =
 \langle\Phi(t)|\Phi(t)\rangle ,
\end{equation}
up to the normalization convention adopted for the propagated wave packet. We use the full time-dependent wave packet without subtracting the initial component, so that the internal and outward-developing parts appear in the same density. In the $T$-type coordinates, $r=r_{\alpha\alpha}$ and $R=R_{n-(\alpha\alpha)}$; in the $Y$-type coordinates, $r=r_{\alpha n}$ and $R=R_{\alpha-(\alpha n)}$. The former is suited to identifying the development of an $n+(\alpha\alpha)$ configuration, whereas the latter exposes correlations in the $\alpha n$ subsystem. For the $Y$-type density, we also isolate the $p_{3/2}$-like $\alpha n$ component to examine the internal $\alpha n$ correlation.

\section{Effective-range description of a virtual state}

The use of an effective-range form is motivated by our previous studies, in which the low-energy $1/2^+$ strength in the $\alpha+\alpha+n$ model was found to be dominated by the $^8$Be+$n$ continuum, with no isolated $1/2^+$ resonance pole identified~\cite{Odsuren2015,Kikuchi2016}. This behavior suggests a strong $^8$Be+$n$ virtual-state character. We therefore use the $s$-wave $^8$Be+$n$ effective-range form as a reference for examining the late-time dynamics of the present three-body wave packet. For an $s$-wave $^8$Be+$n$ virtual state, the effective-range scattering amplitude is
\begin{equation}
f(k)=\left[-\frac{1}{a}+\frac{1}{2}r_0k^2-ik\right]^{-1},
\end{equation}
where $k$ is the $^8$Be+$n$ relative wave number, and $a$ and $r_0$ are the $s$-wave scattering length and effective range, respectively. The relative energy measured from the $^8$Be+$n$ threshold is
\begin{equation}
 E=\frac{\hbar^2k^2}{2\mu},
\end{equation}
with reduced mass $\mu$. For a virtual state, the pole closest to threshold is written as
\begin{equation}
 k=-i\kappa_0,
 \qquad
 \bar E=\frac{\hbar^2\kappa_0^2}{2\mu},
\end{equation}
with $\kappa_0>0$. The second root of the effective-range denominator lies at $k=i\kappa_1$ and defines
\begin{equation}
 E_1=\frac{\hbar^2\kappa_1^2}{2\mu}.
\end{equation}
Accordingly, the denominator of $f(k)$ can be factorized as
\begin{equation}
 -\frac{1}{a}+\frac{1}{2}r_0k^2-ik
 =\frac{r_0}{2}(k+i\kappa_0)(k-i\kappa_1),
\label{eq:ERfactor}
\end{equation}
which gives
\begin{equation}
 r_0=\frac{2}{\kappa_1-\kappa_0},
 \qquad
 a=-\frac{\kappa_1-\kappa_0}{\kappa_0\kappa_1}.
\label{eq:ERparameters}
\end{equation}
Thus, the positive energy scales $\bar E$ and $E_1$ provide an equivalent parametrization of the scattering length and effective range. Together with the $s$-wave threshold factor $k\propto\sqrt{E}$, they determine the dominant low-energy dependence of the strength. Following Efros and Bang~\cite{EfrosBang1999}, we denote the corresponding effective-range spectral weight by $w_{\rm ER}(E)$ and write it as
\begin{equation}
w_{\rm ER}(E)\propto
\frac{2\sqrt{E\bar E}}{E+\bar E}
\frac{\bar E+E_1}{E+E_1},
\qquad E\ge0,
\label{eq:ER}
\end{equation}
where $-\bar E$ corresponds to the virtual-state pole energy within the effective-range parametrization. The two factors in the denominator follow directly from the pole factors in Eq.~(\ref{eq:ERfactor}), while the numerator contains the $s$-wave threshold dependence. Apart from a smooth process-dependent factor, Eq.~(\ref{eq:ER}) therefore gives the dominant low-energy dependence of the virtual-state spectrum~\cite{EfrosBang1999}. We use this form as a reference spectral distribution for the calculated $1/2^+$ decay, not as a resonance parametrization. We note that the effective-range form of Ref.~\cite{EfrosBang1999} is formulated for an effective two-body $^8$Be+$n$ system, whereas the present calculation retains the full $\alpha+\alpha+n$ dynamics. The fitted parameters are therefore used here as effective parameters characterizing the low-energy three-body spectrum and are not identified directly with two-body $^8$Be--$n$ scattering parameters.

The associated survival amplitude is
\begin{equation}
A_{\rm ER}(t)=\mathcal N^{-1}
\int_0^\infty w_{\rm ER}(E)e^{-iEt/\hbar}\,dE,
\label{eq:ERFourier}
\end{equation}
where $\mathcal N=\int_0^\infty w_{\rm ER}(E)dE$, and $P_{\rm ER}(t)=|A_{\rm ER}(t)|^2$. Unlike a Breit--Wigner distribution associated with an isolated pole, the effective-range spectrum retains the explicit threshold energy dependence. Its Fourier transform therefore provides a direct time-domain reference for nonexponential threshold decay. We compare $P_{\rm ER}(t)$ with the late-time three-body survival probability. This comparison is complementary to the fit-window analysis of $\Gamma_{\rm eff}$: the latter tests whether a stationary decay width emerges, whereas the former tests whether the calculated time dependence is compatible with the virtual-state spectral form of Eq.~(\ref{eq:ER}). The parameters $\bar E$ and $E_1$ are determined from the survival-probability comparison under the ordering $E_1>\bar E$.

\section{Results and discussion}

\subsection{$5/2^+$ resonance benchmark}

\begin{figure}[t]
\centering
\includegraphics[width=\linewidth]{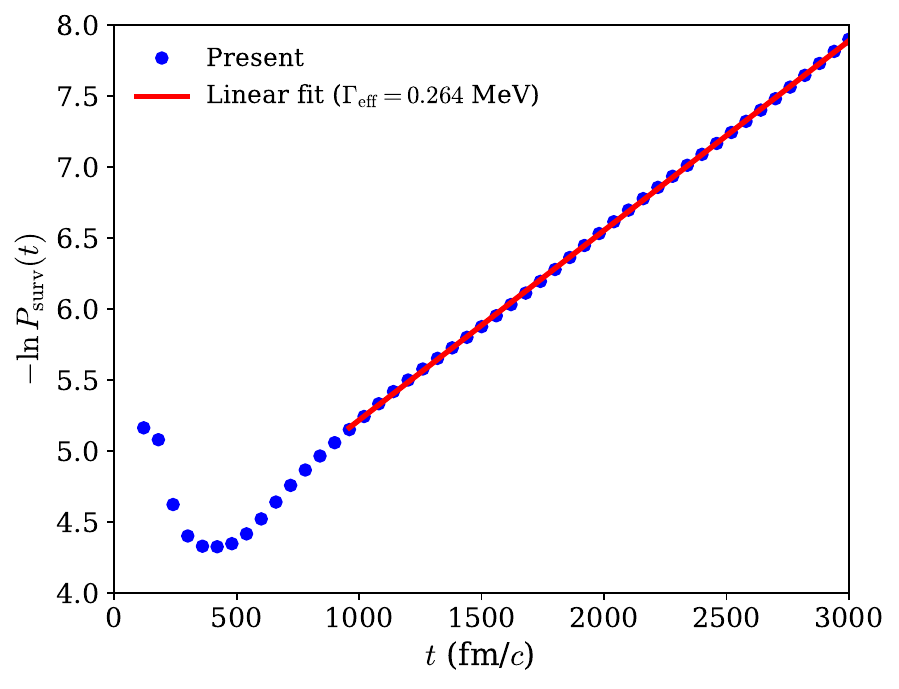}
\caption{Logarithmic survival probability for the $5/2^+$ channel. Blue points show the calculated values, and the red solid line shows the linear fit used to extract $\Gamma_{\rm eff}$.}
\label{fig:fivehalf}
\end{figure}

We first use the $5/2^+$ channel as a reference for decay governed by a well-established resonance. Figure~\ref{fig:fivehalf} shows $-\ln P_{\rm surv}(t)$. After the initial transient, the time dependence becomes nearly linear. A fit over $960$--$3000$ fm/$c$ gives $\Gamma_{\rm eff}(5/2^+)=0.264$ MeV. The fit-window dependence is weak compared with that of the $1/2^+$ channel, with $\Gamma_{\rm eff}$ ranging from approximately $0.264$ to $0.273$ MeV over the windows considered below.

The complex-scaled spectrum contains an isolated $5/2^+$ pole at $E_{\rm CSM}=1.394-i\,0.142$ MeV, corresponding to $\Gamma_{\rm CSM}=0.285$ MeV. The relatively weak fit-window dependence of $\Gamma_{\rm eff}$ and its proximity to the independently determined pole width make the $5/2^+$ channel a useful reference for approximately exponential resonance decay within the same three-body framework and $E1$ preparation procedure.

\subsection{Decay dynamics of the $1/2^+$ state}

We next turn to the $1/2^+$ channel. In contrast to the $5/2^+$ benchmark, $-\ln P_{\rm surv}(t)$ does not approach a single linear behavior over the late-time interval considered here. To quantify this nonstationarity, we vary the lower end of the fitting window from $t_{\min}=1000$ to $2500$ fm/$c$ in steps of $250$ fm/$c$, while fixing $t_{\max}=3200$ fm/$c$. The extracted width decreases systematically from $0.134$ MeV at $t_{\min}=1000$ fm/$c$ to $0.103$ MeV at $t_{\min}=2500$ fm/$c$.

\begin{figure}[t]
\centering
\includegraphics[width=\linewidth]{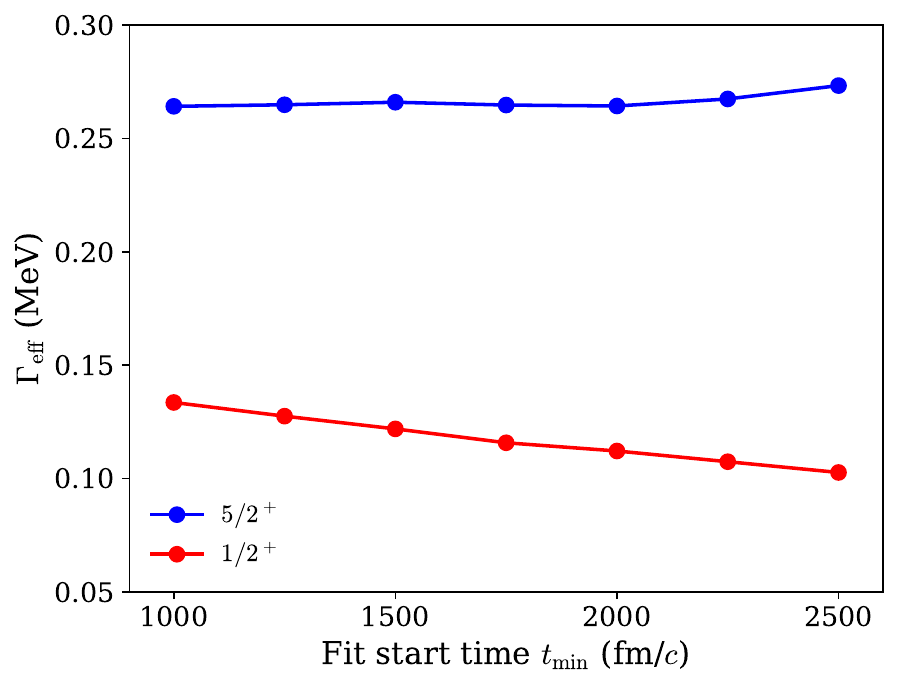}
\caption{Fit-window dependence of $\Gamma_{\rm eff}$. Blue and red points show the $5/2^+$ and $1/2^+$ channels, respectively. The lower limit is varied over $t_{\min}=1000$--$2500$ fm/$c$ in steps of $250$ fm/$c$, while $t_{\max}=3200$ fm/$c$ is fixed.}
\label{fig:window}
\end{figure}

Figure~\ref{fig:window} compares this behavior with the $5/2^+$ benchmark. Whereas the $5/2^+$ width shows only weak dependence on the fitting window, the $1/2^+$ values decrease monotonically as the fit is shifted to later times. Thus, within the time interval considered, the $1/2^+$ decay cannot be characterized by a unique time-independent width.

\begin{figure}[t]
\centering
\includegraphics[width=\linewidth]{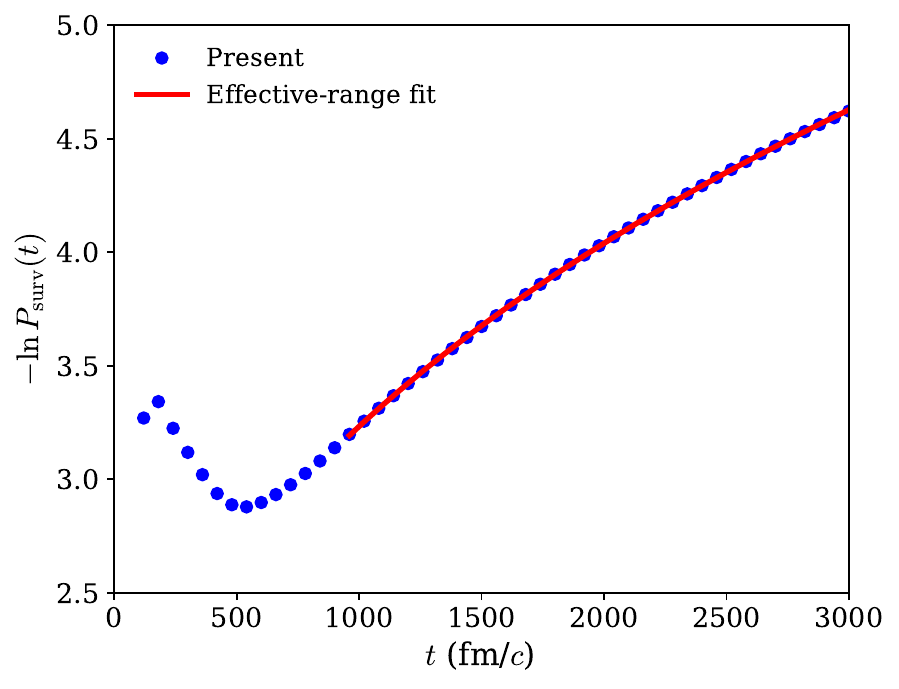}
\caption{Logarithmic survival probability for the $1/2^+$ channel. Blue points show the three-body calculation, and the red solid line shows the result obtained from the effective-range spectral distribution optimized over $960$--$3000$ fm/$c$ under the condition $E_1>\bar E$.}
\label{fig:ERlog}
\end{figure}

We next test whether the effective-range distribution introduced in Sec.~III can reproduce the calculated late-time survival probability. For the fit shown in Fig.~\ref{fig:ERlog}, performed over $960$--$3000$ fm/$c$, the optimized parameters are $\bar E=38.4$ keV and $E_1=161.4$ keV. Within the effective-range parametrization, the former corresponds to a virtual-state pole at $-38.4$ keV. Because the underlying calculation retains the full three-body dynamics, this pole scale should be interpreted as the one associated with the effective-range representation of the late-time spectrum, rather than as a direct two-body pole determination. The Fourier transform of the resulting spectral distribution closely follows the three-body logarithmic survival probability over the fitted interval.

The fit-window dependence and the effective-range comparison provide complementary information. The former shows directly that the calculated decay rate remains nonstationary, whereas the latter demonstrates that the late-time survival probability is compatible with a spectral form containing a near-threshold virtual-state pole. Together, these results favor a virtual-state-like interpretation of the $1/2^+$ decay, although they do not directly determine the pole position on the complex-energy Riemann surface.

\subsection{Spatial evolution of the $1/2^+$ wave packet}

We finally examine the spatial evolution of the $1/2^+$ wave packet. The densities are shown at $t=300$, $600$, and $1200$ fm/$c$. The first two snapshots follow the development away from the initially prepared configuration, while the $1200$ fm/$c$ snapshot lies in the time region where the effective-range description already reproduces the survival probability.

\begin{figure*}[t]
\centering
\includegraphics[width=0.96\textwidth]{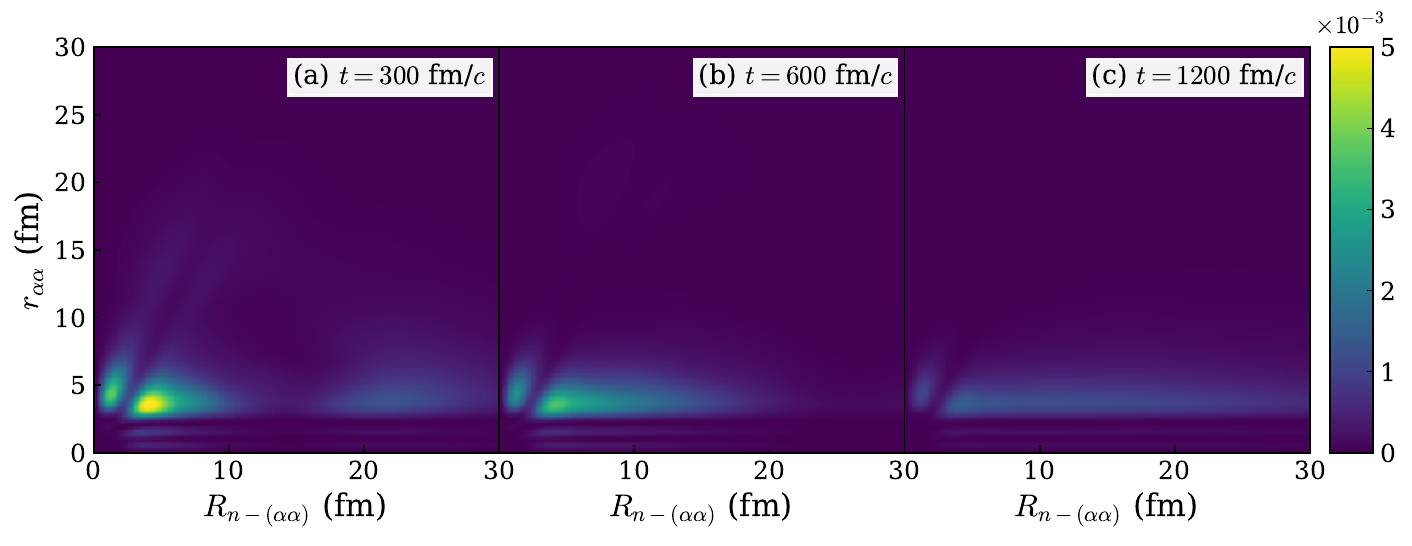}
\caption{Time-dependent $1/2^+$ density in the $T$-type Jacobi coordinates at (a) $t=300$, (b) $600$, and (c) $1200$ fm/$c$. All panels use the same linear color scale.}
\label{fig:Tdensity}
\end{figure*}

\begin{figure*}[t]
\centering
\includegraphics[width=0.96\textwidth]{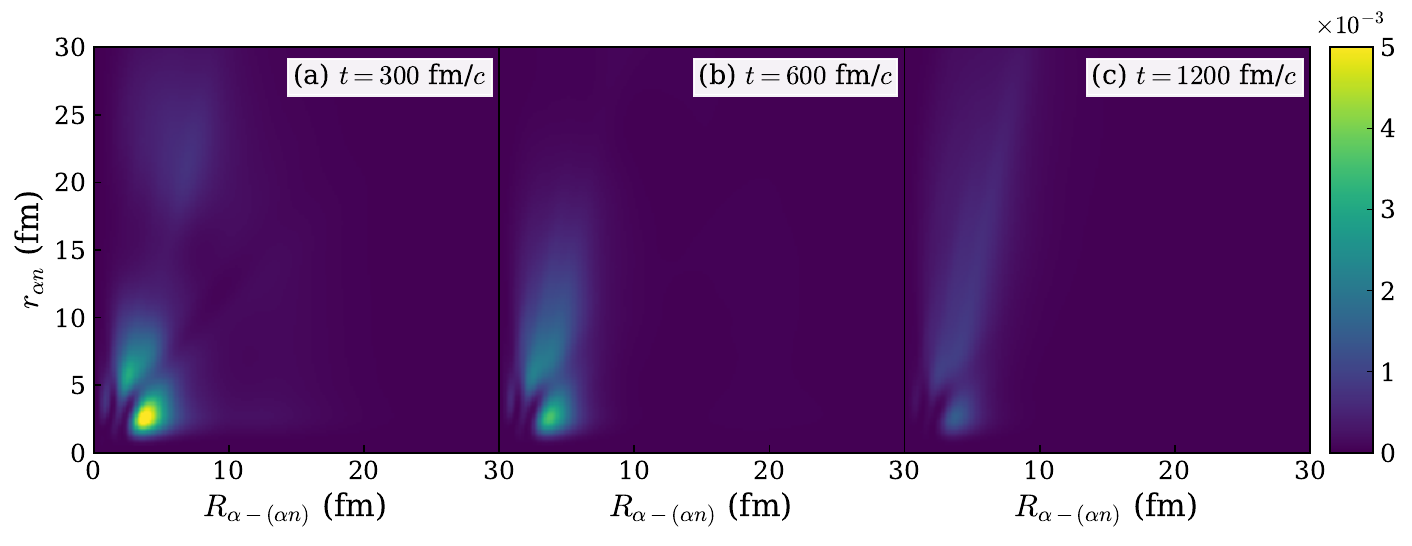}
\caption{Time-dependent $1/2^+$ density in the $Y$-type Jacobi coordinates at (a) $t=300$, (b) $600$, and (c) $1200$ fm/$c$. All panels use the same linear color scale.}
\label{fig:Ydensity}
\end{figure*}

In the $T$-type Jacobi coordinates, Fig.~\ref{fig:Tdensity} shows that the density extends predominantly toward increasing $R_{n-(\alpha\alpha)}$ while the $\alpha\alpha$ separation remains concentrated in the few-femtometer region. The outward-propagating component therefore has predominantly an $n+(\alpha\alpha)$ geometry, naturally associated in the present model with an $^8$Be+$n$-like asymptotic configuration.

The complementary $Y$-type density in Fig.~\ref{fig:Ydensity} retains a localized structure around $r_{\alpha n}\simeq2$--$3$ fm. No comparably clear branch develops toward large $R_{\alpha-(\alpha n)}$ while maintaining a compact $\alpha n$ separation. The compact $\alpha n$ structure is therefore interpreted as an internal correlation rather than as evidence for a dominant asymptotic $^5$He+$\alpha$ decay path.

\begin{figure}[t]
\centering
\includegraphics[width=0.78\columnwidth]{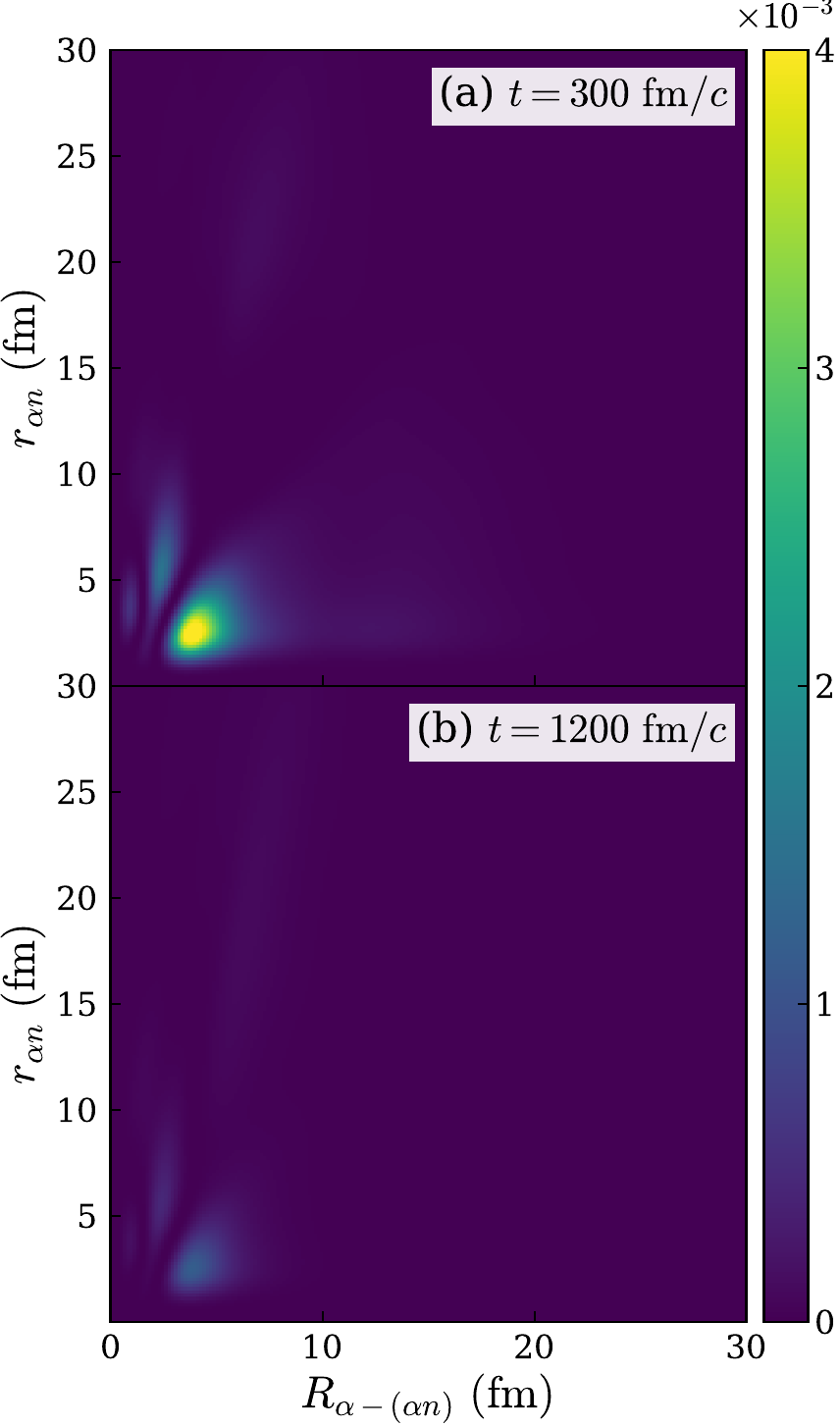}
\caption{Density of the $p_{3/2}$-like $\alpha n$ component in the $Y$-type Jacobi coordinates at (a) $t=300$ and (b) $1200$ fm/$c$. Both panels use the same linear color scale.}
\label{fig:p32density}
\end{figure}

To characterize this internal component, we further decompose the $Y$-type density according to the angular momentum of the $\alpha n$ subsystem. Figure~\ref{fig:p32density} shows the $p_{3/2}$-like component at $t=300$ and $1200$ fm/$c$. A localized peak at $r_{\alpha n}\simeq2$--$3$ fm is already present at $300$ fm/$c$ and persists at $1200$ fm/$c$.

The persistence of this localized $p_{3/2}$-like component is consistent with a $^5$He$(3/2^-)+\alpha$-like configuration in the internal region. Its weak extension toward large $R_{\alpha-(\alpha n)}$ indicates that this component does not develop into a prominent asymptotic $^5$He+$\alpha$ decay path. The outward evolution is instead dominated by the $^8$Be+$n$-like geometry identified in the $T$-type representation.

The coexistence of a compact $^5$He+$\alpha$-like internal structure and an $^8$Be+$n$-like asymptotic configuration is qualitatively compatible with the structural evolution discussed by Garrido, Fedorov, and Jensen~\cite{Garrido2010}. The density evolution alone, however, neither establishes a specific rearrangement mechanism nor demonstrates an isolated three-body resonance. Together with the nonstationary survival dynamics and the agreement with the effective-range description, the density results show that this change in spatial configuration is compatible with virtual-state-like threshold behavior. The spatial and temporal analyses thus provide complementary constraints on the nature of the $1/2^+$ continuum.

\section{Summary}

We have investigated the real-time decay dynamics of low-lying positive-parity states of $^9$Be following $E1$ excitation of the $3/2^-$ ground state in an $\alpha+\alpha+n$ three-body model. The $5/2^+$ state provides a reference for resonance decay: its late-time survival probability approaches an approximately exponential behavior, yielding $\Gamma_{\rm eff}\simeq0.264$ MeV, close to the isolated complex-scaled pole width $\Gamma_{\rm CSM}=0.285$ MeV.

The $1/2^+$ state exhibits qualitatively different behavior. Its effective width decreases systematically as the fitting interval is shifted to later times, and no stationary decay width is obtained within the time range considered. At the same time, the late-time survival probability is well described by an $s$-wave effective-range spectral distribution containing a near-threshold virtual-state pole. Since the present calculation retains the full $\alpha+\alpha+n$ dynamics, the fitted effective-range parameters characterize the low-energy three-body spectral distribution and should not be identified directly with those of a pure $^8$Be+$n$ two-body system. The calculated decay therefore cannot be characterized by a single resonance width and is consistent with virtual-state-like threshold dynamics, although the present analysis does not directly determine the associated pole structure on the complex-energy Riemann surface.

The time-dependent densities provide complementary information on the three-body structure. The outward evolution is dominated by an $^8$Be+$n$-like configuration, while the internal region retains a localized $p_{3/2}$ $\alpha n$ component consistent with a $^5$He$(3/2^-)+\alpha$-like configuration. This distinction between the internal and asymptotic structures is qualitatively compatible with the structural evolution proposed in the three-body-resonance picture. The decay dynamics, however, remain nonstationary and consistent with the effective-range description of a near-threshold virtual state. Taken together, the temporal and spatial analyses provide a dynamical characterization of the $^9$Be $1/2^+$ continuum beyond a description based on a single decay width.

\begin{acknowledgments}
The work shown in this paper was supported by JSPS KAKENHI Grants No. JP26K07092, No. JP25H01268, and JST ERATO Grant No. JPMJER2304, Japan.
\end{acknowledgments}

\end{document}